\documentclass{article}
\usepackage{latexsym}

\begin{document}

\title{A New Look at the Higgs-Kibble Model}
\author{Othmar Steinmann\\Fakult\"at f\"ur Physik\\Universit\"at Bielefeld\\33501 Bielefeld, Germany}
\date{}
\maketitle \vspace{-6mm}
 {\em Dedicated to Wolfhart Zimmermann on the occasion of his $80^{th}$ birthday}

\vspace{2mm}

\begin{abstract}
An elementary perturbative method of handling the Higgs-Kibble
models and deriving their relevant properties, is described. It is
based on Wightman field theory and avoids some of the mathematical
weaknesses of the standard treatments. The method is exemplified
by the abelian case. Its extension to the non-abelian gauge group
$SU_2$ is shortly discussed in the last section.
\end{abstract}

\section{Introduction}

The spontaneous breaking of gauge invariance as described by the
Higgs-Kibble model (henceforth HKM) is an essential ingredient of
the electro-weak part of the standard model of elementary particle
physics. In the present work  we will report on a new, rather
elementary, method of deriving the properties of the model, in
particular its renormalizability (or lack thereof, see Sect.6).

Our method is entirely perturbative, it consists predominantly in
studying the properties of so-called `sector graphs', a simple
generalization of Feynman graphs. But the corresponding graph
rules are derived in an unconventional way. We do not use path
integrals, a not entirely convincing method because of the lack of
a solid mathematical underpinning. Nor do we use the canonical
formalism with its own weak points, like the dubious status of the
canonical commutation relations on account of the non-existence of
interacting fields at a sharp time, and the need for introducing
and handling constraints. Instead we work with an adaptation of
the method introduced in \cite{St1} for QED, where many details
are found beyond what can be reported here.

We will concentrate on the case of the abelian HKM. The extension
of our method, and  of its results, to the non-abelian case will,
however, be briefly described in the last section. Also, we will
work throughout at a formal, non-renormalized level, only getting
as far as obtaining the power-counting behavior necessary for
establishing renormalizability. This sticking to non-renormalized
expressions is not as bad as it sounds. We propose that the theory
be renormalized by Zimmermann's method (known as BPHZ, see
\cite{Zi}), which consists in subtracting not integrals, but the
integrands of the Feynman graphs or, in our case, the sector
graphs. And the cancellations between graphs that we need to
establish for obtaining renormalizability, also happen for the
integrands. Therefore we need only talk about the well defined
integrands, and the divergence (before renormalization) of the
integration over them need not unduly bother us.

\section{The Model}

Let us start with a brief reminder of the definition of the
HKM.\footnote{The standard lore about spontaneous symmetry
breaking can be found e.g. in \cite{IZ},\cite{We}.} The abelian
HKM is a relativistic field theory containing a complex scalar
field $\Phi(x)$ and a real vector field $A_{\mu}(x)$. Its dynamics
is specified by the Lagrangian
\begin{eqnarray}
L &=& -\,\frac{1}{4}\,F_{\alpha\beta}F^{\alpha\beta} +
(\partial_{\alpha}-igA_{\alpha})\,\Phi^*\,(\partial^{\alpha}+igA^{\alpha})\,\Phi
\nonumber \\
 & & + \mu^2\,\Phi^*\Phi - \lambda\,(\Phi^*\Phi)^2     \label{1}
  \end{eqnarray}
with
\begin{equation}
F_{\alpha\beta}(x) = \partial_{\alpha}A_{\beta}(x) -
\partial_{\beta}A_{\alpha}(x) .   \label{2}
\end{equation}
$g,\,\lambda,\,\mu$, are positive real numbers. An important
feature of this Lagrangian is the `wrong' sign of the mass term
$\mu^2\Phi^*\Phi$. $L$ is invariant under the gauge transformation
\begin{equation} \Phi(x)\Rightarrow\exp[i\,g\,\vartheta(x)]\,\Phi(x),\
 \ \ A_{\mu}(x) \Rightarrow
A_{\mu}(x)-\partial_{\mu}\vartheta(x)\label{3}
\end{equation}
for real
functions $\vartheta$.

Because of the unconventional mass term, the field equations
derived from $L$ possess the non-trivial classical solution of
lowest energy
\begin{equation}
\Phi(x)=\Phi^*(x)= \frac{v}{\sqrt{2}}\ ,\ \ \ \ A_{\mu}(x) = 0
\label{4}
\end{equation}
with
\begin{equation}
v = \mu/\sqrt{\lambda}\:>0\,.     \label{5}
\end{equation}
Other solutions of the same lowest energy are generated from
(\ref{4}) by applying gauge transformations (\ref{3}). But they
are of no concern to us.

\medskip

Our perturbative quantum solution consists essentially in a
quantum expansion around the real solution (\ref{4}). We make the
ansatz
\begin{equation}
\Phi(x)=\frac{1}{\sqrt{2}}\,\big(v+R(x)+i\,I(x)\big)\,,  \label{6}
\end{equation}
where $R$ and $I$ are two real fields. Henceforth we treat
$A_{\mu},\,R,\,I,$ as the fundamental fields of the model, while
$\Phi$ is forgotten. With these new fields the solution (\ref{4})
takes the trivial form
\begin{equation}
A_{\mu}=R=I=0\,.          \label{7}
\end{equation}
The gauge transformation (\ref{3}) can be transcribed into the new
fields. We will not write the result down since we are not going
to use it, apart from the important fact that $F_{\alpha\beta}$
and the `Higgs field'
\begin{equation}
\Psi(x) = R(x) + \frac{1}{2v}\big[R^2(x)+I^2(x)\big]     \label{8}
\end{equation}
are gauge invariant.

The Lagrangian (\ref{1}) can also be transcribed into the new
fields. It takes the form
\begin{equation}
L = L_2 +L_3 + L_4\ ,                           \label{9}
\end{equation}
where $L_i$ collects the terms of order $i$ in the fields. A
constant term $L_0$ has been dropped as being immaterial.
Furthermore, we replace $\lambda,\,\mu,$ as parameters of the
theory by
\begin{equation}
 m = g\,v = \frac{g\,\mu}{\sqrt{\lambda}}\,,\ \ \ \ M =
 \sqrt{2}\,\mu\ ,\ \ \ \ \                 \label{10}
\end{equation}
which denote the masses of the gauge boson and the Higgs particle
respectively. They are therefore measurable quantities (barring
the need for renormalization), and they will as usual be kept
fixed. Perturbation theory amounts then to a power series
expansion in the remaining coupling constant $g$. The $L_i$ read
\begin{eqnarray}
L_2& =& -\frac{1}{4}\,F_{\alpha\beta}\,F^{\alpha\beta} +
\frac{m^2}{2}\,A_{\alpha}A^{\alpha} +
m\,A^{\alpha}\partial_{\alpha}I    \nonumber \\
 & & + \frac{1}{2}\,(\partial_{\alpha}R\,\partial^{\alpha}\!R - M^2R^2)
 + \frac{1}{2}\,\partial_{\alpha}I\,\partial^{\alpha}\!I\ ,
 \label{11}
\end{eqnarray}
\begin{eqnarray}
 L_3 &=& - g\,A^{\alpha}I\partial_{\alpha}R +
g\,A^{\alpha}\partial_{\alpha}I\,R + g\,m\,A_{\alpha}A^{\alpha}R
\nonumber \\
 & & -\,\frac{gM^2}{2m}\,R^3 - \frac{gM^2}{2m}\,R\,I^2\ ,
 \label{12}
\end{eqnarray}
\begin{eqnarray}
L_4 &=& \frac{1}{2}\,g^2A_{\alpha}A^{\alpha}\,R^2 +
\frac{1}{2}\,g^2 A_{\alpha}A^{\alpha}\,I^2 \nonumber \\
 & & -\frac{g^2M^2}{8m^2}\,R^4 - \frac{g^2M^2}{8m^2}\,I^4 -
 \frac{g^2M^2}{4m^2}\,R^2I^2\ .        \label{13}
\end{eqnarray}
$L_2$ will be responsible for the propagators of our graph rules,
$L_{int}=L_3+L_4$ for the vertices. The Higgs field takes the form
\begin{equation}
\Psi(x) = R(x) + \frac{g}{2m}\,[R^2(x) + I^2(x)]\ .  \label{14}
\end{equation}

In our method the dynamics is embodied in the field equations
rather than in the Lagrangian. They take the form
\begin{eqnarray}
(\Box + m^2)A^{\mu} - \partial^{\mu}\partial_{\nu}A^{\nu} +
m\,\partial^{\mu}\!I &=& - \frac{\delta L_{int}}{\delta A_{\mu}}\
=:\ \mathcal{R}^{\mu}(x)\ ,               \label{15} \\
-\Box I - m\,\partial_{\nu}A^{\nu} &=& -\frac{\delta
L_{int}}{\delta
I}\  =:\  \mathcal{R}_I(x)\ ,             \label{16}  \\
 -(\Box +
M^2)\,R &=& - \frac{\delta L_{int}}{\delta R} \ =:\
\mathcal{R}_R(x)  \ .                        \label{17}
\end{eqnarray}

As a consequence of the gauge freedom of the theory we note the
following fact. Applying the derivation $\partial_{\mu}$ to the
left-hand side of (\ref{15}) we obtain the left-hand side of
(\ref{16}), up to a constant factor. The equations
(\ref{15})--(\ref{17}) can therefore possess solutions only if the
{\em consistency condition}
\begin{equation}
\mathcal{F} := \partial_{\mu}\mathcal{R}^{\mu} + m\,\mathcal{R}_I
= 0                                         \label{18}
\end{equation}
is satisfied. That this condition is satisfied in our case is
essentially a consequence of the field equations having been
derived from a Lagrangian. It must, however, be noted that in an
explicit verification the field equations must be used. This
verification runs as follows. As contribution of $L_3$ to
$\mathcal{F}$ we find
\begin{equation}
\mathcal{F}_3 = g\,I\,(\Box + M^2)R - g\,R\,(\Box I +
m\,\partial_{\mu}A^{\mu})\ .          \label{19}
\end{equation}
Using the field equations (\ref{16}) and (\ref{17}) this becomes a
polynomial of order 3 in the fields which exactly cancels the
$L_4$-contribution
\begin{eqnarray}
\mathcal{F}_4 &=& - g^2R^2\partial_{\mu}A^{\mu} -
2\,g^2R\,\partial_{\mu}R\,A^{\mu} - g^2I^2\partial_{\mu}A^{\mu} -
2g^2I\partial_{\mu}I\,A^{\mu} \nonumber \\ & &
-m\,g^2I\,A_{\mu}A^{\mu} + \frac{g^2M^2}{2m}\,I^3 +
\frac{g^2M^2}{2m}\,R^2I\ .               \label{20}
\end{eqnarray}
This looks at first like a consistency check rather than a proof.
It is, however, perfectly acceptable as a proof in perturbation
 theory.

\medskip

We will not endeavor to give a general definition of what we
understand under a particular gauge of this model. But the
following statement is essential. {\em A quantum field theory
claiming to be the HKM in a particular gauge must satisfy the
field equations (\ref{15})--(\ref{17}).} In the following section
a particular class of gauges will be constructed.

\section{Wightman Gauges}

Under `Wightman gauges' we understand a class of quantum field
theories solving the field equations (\ref{15})--(\ref{17}), and
moreover satisfying the Wightman axioms (see \cite{SW}), i.e.
Poincar\'e covariance, locality, spectral condition, existence of
a vacuum, and the cluster property, with the possible exception of
positivity. This last condition can in general not be expected to
hold in a gauge theory. Two special cases of Wightman gauges will
be of particular interest to us. The first is the `unitary' or
`physical' gauge, which allows to specify the physical content of
the theory. And the second is the `renormalization' gauge, which
is particularly suited for establishing the renormalizability of
the physically relevant part of the model.

Our method consists essentially in a recursive solution of the
field equations. But the fundamental objects of the approach are
the Wightman functions (W-functions), not the field operators
themselves, and also not the Green's functions of the conventional
methods. The W-functions are the vacuum expectation values of
ordinary (not time ordered) products of field operators. According
to Wightman's reconstruction theorem \cite{SW} the theory is fully
determined by these W-functions.\footnote{Positivity of the scalar
product is not necessary for the validity of the reconstruction
theorem (see Sect.~4.2 of \cite{St1}).} The field equations
applied to any factor in a W-function produce a set of
differential equations for these functions. And  this set of
differential equations  we solve recursively.

The resulting expression for a given function
$(\Omega,\,\varphi_1(x_1)\cdots\varphi_n(x_n)\,\Omega)$,
$\varphi_i$ any of the fundamental fields $A_{\mu},\:I,\:R$, in a
given order $g^{\sigma}$ of perturbation theory can be written as
a sum over generalized Feynman graphs called `sector graphs'. A
sector graph looks at first just like an ordinary Feynman graph
not containing any vacuum-vacuum subgraphs\footnote{These
subgraphs do not occur in our formulation because we work in the
Heisenberg picture. The vacuum graphs are an artifact of the
interaction picture.}. But its vertices are then partitioned into
non-overlapping subsets called `sectors', in such a way that each
sector contains at most one external point corresponding to one of
the fields in $W$. Lines connecting vertices (including the
external points) in the same sector belong to this sector and are
called `sector lines'. Lines connecting points in different
sectors are called `cross lines'. The sectors can be of two types,
$T^+$ or $T^-$. They are numbered such that the sector containing
the external point belonging to $\varphi_i$ carries the number
$i$. It is convenient to alternate the corresponding sectors:
sectors with an odd number are $T^-$, those with an even number
$T^+$, or vice versa. In this case there occur no sectors without
external points.

The internal vertices correspond as usual to the terms in
$L_{int}$ as listed in (\ref{12}), (\ref{13}). Their vertex
factors are also the conventional ones in $T^+$ sectors, their
complex conjugates in $T^-$ sectors. E.g. the last term in
(\ref{12}) produces a vertex with one $R$-line and two $I$-lines
joining, and with the vertex factor $\mp i\,g\,m^{-1}M^2$ in
$T^{\pm}$ sectors. Note that the $L_4$-vertices are of second
order in $g$.  A cross line joining the vertex with variable $u$
in sector $i$ to the vertex $v$ in sector $j$, $j>i$, carries the
`cross propagator'
\begin{equation}
w_{ab}(u-v) = \langle\varphi_a(u),\varphi_b(v)\rangle_0\ ,
\label{21}
\end{equation}
where $\langle\varphi_a\,\varphi_b\rangle_0$ is a free 2-point
function (to be specified below), and the indices $a,\,b$, signify
the field types of the ends of the line in question. A sector line
connecting the vertices $u$ and $v$ in a $T^{\pm}$ sector carries
as propagator the time ordered or anti-time ordered function
$\tau^{\pm}_{ab}(u-v)$ corresponding to the $w_{ab}$ of
(\ref{21}).

With the rules given as yet there holds the {\em Ostendorf
theorem} \cite{Os},\,\cite{St2},\,\cite{St1}, stating that the so
defined functions $W_{\sigma}$ satisfy all Wightman properties
with the possible exception of positivity. Hence these rules
define a, slightly generalized, Wightman theory.

 \medskip

But we still must satisfy the requirement that these $W$ solve the
interacting field equations (\ref{15})--(\ref{17}). This problem
is easier to handle in $p$-space. Therefore we will from now on
mainly work in this space, with the Fourier transforms of
(\ref{15})--(\ref{17}). That these equations are satisfied in
$0^{th}$ order in $g$ is guaranteed by the condition that the
$w_{ab}$ solve the free field equations. For the following we need
to know the $w_{ab}$ more explicitly. In $p$-space we have
\begin{equation}
\langle\varphi_a(p)\,\varphi_b(q)\rangle_0 =
w_{ab}(p)\,\delta^4(p+q)\ ,                   \label{22}
\end{equation}
where this new $w_{ab}$ is the Fourier transform of the $w_{ab}$
in (\ref{21}). The most general solution of the free field
equations satisfying all Wightman properties, in particular
covariance and locality, is easily found to be
\begin{equation}
\left. \begin{array}{r@{\;=\;}l} w_{\mu\nu}(p) &
-\omega\,\big(g_{\mu\nu}-\frac{p_{\mu}\,p_{\nu}}{m^2}\big)\,\delta^m_+(p)
+ \frac{1}{m}\,p_{\mu}p_{\nu}\,T(p) \\
w_{II}(p) & m\,T(p) \\
w_{\mu I}(p) & -w_{I\mu}(p) = i\,p_{\mu}\,T(p) \\
w_{RR}(p) & \alpha\,\delta^M_+(p) \\
w_{RI}(p) & \beta\,\delta^M_+(p)\ ,\qquad w_{R\mu}(p) =
-i\,\frac{\beta}{m}\,p_{\mu}\,\delta^M_+(p)\ ,
\end{array}
\right\}                        \label{23}
\end{equation}
where $\delta^m_+(p) = \theta(p_0)\,\delta(p^2-m^2)$ is the Dirac
measure for the positive mass shell. $\alpha,\,\beta,\,\omega,$
are as yet undetermined real constants, $T(p) =
\theta(p_0)\,T'(p^2)$ is an arbitrary real invariant function with
support in the forward light cone.

The corresponding (anti-)time ordered functions $\tau^{\pm}(p)$
serving as sector propagators are then also uniquely fixed,
provided we restrict ourselves to $T'$ which tend to 0 for
$p^2\to\infty$, and that we demand that $\tau^{\pm}$ should
increase for $p\to\infty$ as weakly as possible. It turns out that
the resulting W-functions satisfy the interacting field equations,
if the $\tau^{\pm}$ are propagators in the original sense of the
word used in the theory of differential equations. In $p$-space
this means the following. We write the $p$-space form of the field
equations (\ref{15})--(\ref{17}) in matrix notation as
\begin{equation}
C(p)\,\varphi(p) = \mathcal{R}(p)\ .             \label{24}
\end{equation}
Here $C$ is the $6 \times 6$ coefficient matrix
\begin{equation}
C = \left( \begin{array} {ccc} -(p^2-\,m^2)\,\delta^{\mu}_{\nu}+
p^{\mu}p_{\nu} & -i\,m\,p^{\mu} & 0 \\
i\,m\,p_{\nu} & p^2 & 0 \\
0 & 0 & p^2-M^2
\end{array}
\right)\ .                 \label{25}
\end{equation}
The lines are indexed by $\mu,\,I,\,R,$ the rows by $\nu,\,I,\,R,$
where $\mu$ and $\nu$ run over the values $0,\cdots,3$. $\varphi$
is a 6-vector with components ($A^{\nu},\,I,\,R$), $\mathcal{R}$ a
6-vector with components
($\mathcal{R}^{\mu},\,\mathcal{R}_I,\,\mathcal{R}_R$). We call the
$6\times 6$ matrix $\mathcal{P}(p)$ a {\em propagator matrix} if
\begin{equation}
C\,\mathcal{P}\,V = V           \label{26}
\end{equation}
holds for all 6-vectors $V(p)$ satisfying the consistency
condition (\ref{18}):
\begin{equation}
-i\,p_{\mu}V^{\mu} + m\,V_I = 0\ .                \label{27}
\end{equation}
Then
\begin{equation}
\varphi = \mathcal{P}\,\mathcal{R}                  \label{28}
\end{equation}
solves (\ref{24}). Remember that the $I$-line of $C$ is a linear
combination of the $\mu$-lines. Hence $C$ is not invertible and
$\mathcal{P}$ cannot be defined as its inverse. Therefore the
restriction (\ref{27}) is necessary.

The sum over our sector graphs solves the field equations of the
HKM if
\begin{equation}
\mathcal{P}^{\pm}_{ab}(p) = \mp\,2\pi  i\,\tau^{\pm}_{ab}(p)
\label{29}
\end{equation}
constitute a propagator matrix. This is seen by applying the field
equations to the propagators of the external graph lines, using
that the internal ends of these lines correspond to $\mathcal{R}$
vertices (see \cite{St1}, Sect.~9.4, for the QED analogue).
External cross propagators do not contribute because they solve
the free field equations. It turns out that  condition (\ref{29})
fixes two of the free constants in (\ref{23}) to be
\begin{equation}
\omega = \alpha = 1\ ,           \label{30}
\end{equation}
while $\beta$ and the function $T(p)$ are still free.

 \medskip

From our rules for calculating W-functions we can also obtain the
rules for the fully time ordered functions. At our present formal,
non-renormalized, level this is simply done by using the formal
definition of time ordering with the help of step functions. The
result is a representation as a sum of graphs with only one $T^+$
sector containing all external points. The corresponding graph
rules are simply the standard Feynman rules. That the Green's
functions thus defined are indeed the time ordered functions of a
field theory is of course essential for the applicability of the
LSZ reduction formula for the calculation of the $S$-matrix.

We will also have occasion to consider functions of the form
\[ \Big(\Omega,\,T^-\big(\varphi_1(x_1)\cdots\varphi_n(x_n)\big)\,
 T^+\big(\psi_1(y_1)\cdots\psi_m(y_m)\big)\,\Omega\Big) \ ,\]
where $\varphi_i$ or $\psi_j$ stands for any of our fields. These
are given by 2-sector graphs with a $T^-$ sector containing all
external $x_i$ points and a $T^+$ sector containing all $y_j$
points.

\section{The Unitary Gauge}

The unitary gauge, or U-gauge, is defined as the special Wightman
gauge obtained by the choice
\begin{equation}
\beta = T(p) = 0\ .                     \label{31}
\end{equation}
In this gauge we have
\begin{equation}
I = 0\ ,                           \label{32}
\end{equation}
it is simply the gauge specified by the `gauge condition'
(\ref{32})\footnote{The other W-gauges cannot be characterized in
this simple way.}. Hence we are left only with the fields $R$ and
$A_{\mu}$. The surviving non-vanishing cross propagators are
\begin{equation}
w_{\mu\nu}(p) =
-\,(g_{\mu\nu}-m^{-2}p_{\mu}p_{\nu})\,\delta^m_+(p)\ ,\quad
w_{RR}(p) = \delta^M_+(p)\ ,              \label{33}
\end{equation}
and the sector propagators are
\begin{eqnarray}
\tau^{\pm}_{\mu\nu}(p) &=& \mp (i/2\pi)\,(g_{\mu\nu} -
m^{-2}\,p_{\mu}p_{\nu})\,(p^2-m^2 \pm i\epsilon)^{-1} \nonumber \\
\tau^{\pm}_{RR}(p) &=& \pm (i/2\pi)\,(p^2-M^2\pm i\epsilon)^{-1}\
. \label{34}
 \end{eqnarray}

\medskip

The special interest of this gauge rests on the fact that it might
also be called the `physical gauge'. The physically relevant
objects of a quantum field theory are the observables and the
physical states.\footnote{The widely held opinion that the
physical content of the theory is fully described by its S-matrix
is not tenable. The S-matrix relates states at positive infinite
times to states at negative infinite time. But we always measure
at finite times. Therefore the S-matrix is, in fact, not
measurable.} In an experiment we usually measure expectation
values of observables in physical states (meaning states that can
actually be prepared in a laboratory). The physical content of a
gauge theory must be gauge independent. For the observables this
implies that they must be gauge invariant. What it means for
states is less easy to characterize. But in the HKM the state
space $\mathcal{V}_U$ of the U-gauge is the obvious candidate for
the role as physical state space. This claim rests on two facts.
First, the cross propagators are positive. For the $w_{\mu\nu}$
this means more exactly that they form a positive matrix. This
implies that our graph rules define on $\mathcal{V}_U$ a positive
scalar product\footnote{A formal power series $Q(g)$ is said to be
positive if there exist formal power series $S_i(g)$ such that
$Q(g) = \sum_iS_i(g)^*\,S_i(g)$\,.}, a necessary requirement for a
physical state space.

The second vital point is the following. At first, $\mathcal{V}_U$
is generated from the vacuum state $\Omega$ by applying to it
polynomials in the fields $R,\,A_{\mu}$, properly integrated over
sufficiently smooth test functions. But it turns out that the same
state space is also created out of $\Omega$ by applying
polynomials in the gauge invariant fields $F_{\alpha\beta}$ and
$\Psi$. This is so because $A_{\mu}$ and $R$ can be expressed as
functions of $F_{\alpha\beta}$ and $\Psi$. We see this as follows.
The definition (\ref{14}) of $\Psi$ becomes in the U-gauge
\begin{equation}
\Psi = R + \frac{g}{2m}\,R^2\ .          \label{35}
\end{equation}
This equation can in principle be solved for $R$. Of course,
square roots of operators are not easy to deal with. But we work
in perturbation theory, and here there is no problem. Expand $R$
in a power series:
\begin{equation}
R = \sum^{\infty}_{\sigma=0} R_{\sigma}g^{\sigma}\ ,  \label{36}
\end{equation}
and similarly for $\Psi$, and insert these expansions into
(\ref{35}). We find in zeroth order
\[ R_0 = \Psi_0\ ,   \]
in first order
\[  R_1 = \Psi_1 - (2m)^{-1}\,{R_0}^2 = \Psi_1 - (2m)^{-1}\,{\Psi_0}^2\ ,\]
and so on. The fact that for increasing $\sigma$ $R_{\sigma}$
becomes a polynomial in $\Psi_{\varrho},\ \varrho\leq\sigma,$ of
indefinitely increasing order need not worry us, because this
expansion will never be used explicitly. Next, from the definition
of $F_{\alpha\beta}$ and the field equation (\ref{15}) we obtain
\begin{equation}
\partial^{\alpha}F_{\alpha\beta}= \Box A_{\beta}-
\partial_{\beta}\partial^{\alpha}\!A_{\alpha} = -m^2 A_{\beta}+
\mathcal{R}_{\beta}(A_{\mu},\,R)\ ,     \label{37}
\end{equation}
or its Fourier transform, hence
\begin{equation}
m^2 A_{\beta} = - \partial^{\alpha}\!F_{\alpha\beta}+
\mathcal{R}_{\beta}(A_{\mu},\,R)\ .      \label{38}
\end{equation}
Since $\mathcal{R}_{\beta}$ contains an explicit factor $g$, this
equation allows again an iterative expansion of $A_{\mu}$ in
polynomials of $F_{\mu\nu}$ and $\Psi$.

Hence $\mathcal{V}_U$ has an explicitly gauge invariant structure,
which fact justifies the claim that it is the physical state space
of the HKM.

 \medskip

In this way we seem to have arrived at a nice, clean,
identification of the physical content of the HKM. There is,
however, a fly in the ointment. The $p_{\mu}p_{\nu}$ term in
(\ref{34}) has a bad behavior at large $p$, leading to
non-renormalizability of the theory in the simple power counting
sense. In increasing orders of perturbation theory the individual
graphs will have an increasingly bad ultraviolet behavior. The
claim that the theory is nevertheless renormalizable amounts to
claiming that these bad UV contributions in individual graphs
cancel in the sum of all graphs contributing to a specific
W-function (or time ordered function) in a given order $\sigma$ of
perturbation theory.

The standard way of handling this problem consists in adding a
so-called `gauge fixing term'
\begin{equation}
-\,\frac{1}{2\,\alpha}\,(\partial_{\mu}A^{\mu})^2    \label{39}
\end{equation}
to the original Lagrangian of the model. The theory thus obtained
is renormalizable in the  sense of power counting. But calling
(\ref{39}) a gauge fixing term is highly misleading. The amended
$\alpha$-Lagrangian does by no means describe a particular gauge
of the HKM. It defines a new, different theory, which does {\em
not} solve the field equations of the HKM. Hence its
renormalizability is of no use to our problem, unless it can be
established to be in some way physically equivalent to the HKM, in
particular to its U-gauge formulation. This necessity does not
quite find sufficient attention in the literature.

In any case, if the claimed cancellations between graphs really
happen, this ought to be provable inside the HKM. This is the task
that we now turn to. There seems to be no easy way to achieve this
purpose in the U-gauge. Therefore we introduce in the next section
another Wightman gauge better suited to the task.

\section{Renormalizability}

Particularly suited for our purpose is the {\em R-gauge} (for
`renormalization gauge') specified by the choice
\begin{equation}
\omega = \alpha = 1\,,\quad \beta = 0\,,\quad T(p)
=-\,\frac{1}{m}\,\delta_+(p)                          \label{40}
\end{equation}
in (\ref{23}), with $\delta_+(p) = \delta_+^0(p)$. The
corresponding $T^+$ propagators are
\begin{equation}
\left. \begin{array}{r@{\;=\;}l}
 \tau^+_{\mu\nu}(p) & \Big(-g_{\mu\nu} +
 \frac{p_{\mu}p_{\nu}}{p^2+i\epsilon}\Big)\,\frac{i}{2\pi(p^2-\,m^2+i\epsilon)}
 \\
 \tau^+_{II}(p) & - \frac{i}{2\pi(p^2+\,i\epsilon)}  \\
\tau^+_{\mu I}(p) & - \tau^+_{I\mu}(p) =
\frac{p_{\mu}}{2\pi\,m\,(p^2+i\epsilon)} \\
\tau^+_{RR}(p) & \frac{i}{2\pi(p^2-M^2+i\epsilon)}
\end{array}
\right\}\ .                  \label{41}
\end{equation}
The $\tau^-$ are obtained from $\tau^+$ by the replacements $(i\to
-i,\ p\to -p)$.

Notice that now the propagator $\tau^+_{\mu\nu}$ has a nice,
renormalizable, large-$p$ behavior, at the price of introducing
the ghost factor$(p^2)^{-1}$. Unfortunately, that does not mean
that the theory has become renormalizable. The bad UV behavior has
merely been shifted to the mixed $A$-$I$ propagators. But here the
desired cancellations are easier to prove than in the U-gauge.

Before attacking this problem we must decide how the physical
content of the model presents itself in the new gauge. Remember
that in the U-gauge the physical state space is generated from the
vacuum by applying polynomials in the gauge invariant fields
$F_{\alpha\beta}$ and $\Psi$. Since the physical content of the
theory must be gauge invariant, the same will be true in the
R-gauge: the physical state space $\mathcal{V}_{ph}$, which is now
a proper subspace of the full state space, is generated from the
vacuum by applying polynomials in $F_{\alpha\beta}$ and $\Psi$.
$\mathcal{V}_{ph}$ can again be reconstructed by the Wightman
reconstruction theorem from the W-functions of these physical
fields only. Only in these `physical' W-functions are we really
interested, hence only for them need we prove renormalizability.

The graph representation of the physical fields is clear. To
obtain an external $F_{\alpha\beta}$ propagator, simply replace
the factor $(-g_{\mu\nu} + p_{\mu}p_{\nu}/p^2)$ of an external
$A_{\mu}$ line by $i\,(p_{\alpha}g_{\beta\nu}
-p_{\beta}g_{\alpha\nu})$, the index $\nu$ belonging to the
adjacent internal vertex. A $\Psi(p)$ factor is represented as a
sum of three terms, an ordinary external $R$-line plus two
external 2-prong vertices representing the composite fields $R^2$
and $I^2$ in (\ref{14}). Both these composite vertices carry the
vertex factor $g/(\sqrt{2\pi}^{\,3}m)$.\footnote{The Fourier
transform of the field $\varphi(x)$ is defined as $\varphi(p)=
(2\pi)^{-5/2} \int dx\,e^{ipx} \varphi(x)$.}

 \medskip

We turn now to the promised proof of the cancellation of UV
dangerous terms. The basic idea is the following. Consider a
$I$--$A_{\mu}$ cross propagator
\[  w_{I\mu}(p) = i\,\frac{p_{\mu}}{m}\,\delta_+(p)\ ,    \]
derived by (\ref{22}) from the free 2-point function $\langle
I(p)\,A_{\mu}(q)\rangle_0$. The end vertex of the corresponding
cross line corresponds to a term in $\mathcal{R}^{\mu}(q)$.
Summing over all these terms we obtain (with $q=-p$)
\begin{equation}
\delta_+(p)\,\big(-i\,\frac{q_{\mu}}{m}\,\mathcal{R}^{\mu}(q)\big)
= - \delta_+(p)\,\mathcal{R}_I(q)           \label{42}
\end{equation}
by the consistency condition (\ref{18}). This means that we can
replace the UV nice vertex sum $\mathcal{R}^{\mu}$ by the equally
UV nice $\mathcal{R}_I$, and the UV bad propagator $w_{I\mu}(p)$
by the UV nice $-w_{II}(p)$! Unfortunately, in this crude form the
argument is incorrect. The $\mathcal{R}^{\mu}(q)$ vertex in
question belongs to, let us say, a $T^+$ sector, which represents
a time ordered function of its external vertices, including the
$\mathcal{R}$ vertex with $\mathcal{R}^{\mu}$ considered a
composite external field. But in $x$-space the propagator factor
$-i\,q_{\mu}$ represents a derivation $\partial_{\mu}$ acting not
only on $\mathcal{R}^{\mu}(x)$ but also on the step functions
occurring in the definition of the $T$-product. Hence we must
expect that the relevant quantity
\[ \frac{-i\,q_{\mu}}{m}\,\tau^+\big(\mathcal{R}^{\mu}(q)\cdots\big) +
 \tau^+\big(\mathcal{R}_I(q)\cdots\big) = \frac{1}{m}\, \tau^+\big(\mathcal{F}(q)\cdots\big)     \]
does not vanish but is given as a sum of contact
terms.\footnote{The explicit form of this relation is known as a
Ward-Takahashi identity.} Luckily it turns out that these contact
terms are not present if the sector in question contains only
gauge invariant external fields. This is established by an
explicit study of the graphs in question. Consider first the case
that the $\mathcal{R}(q)$ vertices are those coming from $L_3$.
Then the $\mathcal{F}_3$ occurring on the right-hand side is the
Fourier transform of the expression (\ref{19}). Consider a
$R$-line with momentum $k$ issuing from the vertex in question.
Its denominator $(k^2-\,M^2)^{-1}$ is cancelled by the numerator
$(k^2-\,M^2)$ coming from the first term in (\ref{19}). Thus this
first term  leads to an amputation of the adjoining $R$-line, and
a corresponding fusion of its two end vertices (internal or
external) into a single vertex with more lines. The second term in
(\ref{19}) produces the same effect on  $I$- and $A$-lines
starting from the $\mathcal{F}$ vertex. In this way we obtain a
considerable number of fused vertices, among which extensive
cancellations occur. And the remaining fused vertices cancel
against the $L_4$ terms in $\mathcal{F}(q)$. The actual
verification of these cancellations is completely elementary but
rather lengthy and tedious on account of the large number of
different vertices to be considered (see (\ref{12}), (\ref{13}),
(\ref{20})). The remarkable thing is, however, that these
cancellations happen locally in the graphs in the immediate
neighborhood of the $q$-end of the cross line in question,
involving only that end vertex and its nearest neighbors, no
matter how large the full sector may be. As a result we can, as
proposed, drop our bad $I$--$A_{\mu}$ cross line and replace it by
the negative of a good $I$--$I$ line. The same argument, now used
for the starting point, applies of course to a $A_{\mu}$--$I$
cross propagator. It may also be replaced by the negative of a
$I$--$I$ propagator. By this we end up with two negative $I$--$I$
propagators for a given position of an appropriate line, plus the
positive $I$--$I$ propagator present from the beginning. The net
effect is that we drop the dangerous mixed cross propagators and
change the sign of the $I$--$I$ cross propagators without changing
our physical W-functions.

In this consideration we have assumed that the internal
propagators in the sectors involved still have the original
R-gauge form, and that the same applies to other cross propagators
possibly involved in the cancellations. But the remarkable and
lucky fact is that the said cancellations also occur if we have
already effected the changes of rules explained above inside the
sectors in question and in some of the cross lines, i.e. if we
have already dropped there the mixed propagators and changed the
signs of the $I$--$I$ propagators. This enables us to prove the
following

 \medskip

{\bf Theorem.}
\begin{it}
If in the graph rules of the R-gauge we omit the mixed
$A_{\mu}$--$I$ lines and change the signs of the $I$--$I$
propagators, then the resulting Wightman functions and related
(partially or fully time ordered) functions of the physical fields
$F_{\alpha\beta},\ \Psi$, remain unchanged.
\end{it}

 \medskip

Notice that the new graph rules arrived at in this way are those
of the case $\alpha=0$ (`Landau gauge') of the conventional
$L_{\alpha}$ approach, thus confirming the perturbative validity
of that approach. These new graph rules are clearly renormalizable
in the sense of power counting. In fact, they are also
renormalizable in the stricter sense that the necessary
subtractions can be fully absorbed into renormalizations  of the
masses $m,\:M$, the coupling constant $g$, and the field
normalizations. But the proof of this is quite involved and lies
outside the scope of the present work.

The proof of the Theorem is inductive with respect to the order
$\sigma$ of perturbation theory. It consists of the following
points.

 \medskip

(1) {\it The theorem is correct for $\sigma\leq 2$}. This is
easily established by explicit calculation.

(2) {\it If the theorem is true for the 2-sector functions
$\big(\Omega,\,T^-(\cdots)\,T^+(\cdots)\,\Omega\big)_{\sigma}$,
then it is true for all $n$-sector functions
$\big(\Omega,\,T_1^{\pm}(\cdots)\:\cdots\:T_n^{\pm}(\cdots)\,\Omega\big)_{\sigma}$,
in particular the W-functions, with the same fields. } This is so
because all these functions are in $x$-space boundary values of
the same analytic function.\footnote{Strictly speaking this is not
true at points where two arguments in the same $T^{\pm}$ factor
coincide. But this is of little concern because it does not happen
in the W-functions, which are the functions of central interest.}
The reason for this is that, first, all permuted W-functions of a
given set of fields are boundary values of a single analytic
function (see \cite{SW}, Theorem 3-6), and that, second, any
$n$-sector function is locally equal to a permuted W-function,
wherever all $x_i^0$ are different and, because of Lorentz
invariance, even where all $x_i$ are different.

  (3) Amputate the considered
functions by multiplying them with $(p^2-m^2) $ for factors
$F_{\alpha\beta}(p)$ , $(p^2-M^2)$ for factors $\Psi(p)$ . {\it
Then the theorem is true for the full functions if it is true for
the amputated ones.} This is so because we know precisely how to
reconstruct the full functions from the amputated ones.

(4) {\it The theorem is true for the amputated 2-sector functions
of order $\sigma$}. This is seen by noticing that in the
corresponding 2-sector graphs both sectors are of orders $\varrho$
with $0<\varrho<\sigma$, so that the inductive hypothesis is
applicable to them: the new rules can be used inside these
sectors. Then the cross propagators linking them can also be
changed to the new form by the arguments related above.

\section{The Non-Abelian Case}

The methods used for the abelian HKM can be extended to the
non-abelian case. In this last section we will briefly describe,
without details,  this extension and its results in the case of
the gauge group $SU_2$.

 \medskip

The fields of the model are a complex 2-vector $\Phi(x)$ with the
scalar components $\phi_1(x),\;\phi_2(x),$ and a triplet
$A^{\mu}_1(x),\cdots,A^{\mu}_3(x),$ of real vector fields. The
Lagrangian is\footnote{We use the summation convention both for
Minkowski indices $\mu,\ldots,$ and group indices $a,\ldots.$}
\begin{eqnarray}
L &=& -\,\frac{1}{4}\,F_{a\!,\,\mu\nu}\,F^{\mu\nu}_a +
\big[\partial_{\mu}-
g\,A_{b,\mu}T_b)\,\Phi\big]^*\big[(\partial^{\mu}-g\,A^{\mu}_c\,T_c)\,\Phi\big]
\nonumber \\
 & & + \mu^2\,\Phi^*\Phi - \lambda\,(\Phi^*\Phi)^2\ .
 \label{43}
\end{eqnarray}
Here
\begin{equation}
F^{\mu\nu}_a = \partial^{\mu}\!A^{\nu}_a -
\partial^{\nu}\!A^{\mu}_a - g\,
\varepsilon_{abc}A^{\mu}_b\,A^{\nu}_c\ ,       \label{44}
\end{equation}
and
\begin{equation}
T_a = - \frac{i}{2}\,\sigma_a\ ,         \label{45}
\end{equation}
$\sigma_a$ the Pauli matrices.

$L$ is invariant under the infinitesimal gauge transformations
\begin{equation}
\left. \begin{array}{r@{\Rightarrow}l} \Phi(x) &
(1+g\,\vartheta_a(x)\,T_a)\,\Phi(x) \\
A^{\mu}_a(x) & A^{\mu}_a +
g\,\varepsilon_{abc}\,\vartheta_b(x)\,A^{\mu}_c(x) +
\partial^{\mu}\vartheta_a(x)
\end{array} \right\}                \label{46}
\end{equation}
for infinitesimal real functions $\vartheta_a$. In contrast to the
abelian case, the field strengths $F_a^{\mu\nu}$ are not gauge
invariant.

The corresponding field equations possess the `vacuum solution'
\begin{equation}
\Phi = \frac{1}{\sqrt{2}}\,\left(\begin{array}{c} v \\ 0
\end{array} \right)\ ,\qquad A_a^{\mu\nu} = 0\ \  \forall a\ ,\qquad
v = \frac{\mu}{\sqrt{\lambda}}\ ,  \label{47}
\end{equation}
which takes over the role of the abelian solution (\ref{4}). The
$\phi_i$ are replaced as fundamental fields by the real scalar
fields $R(x),\;I_a(x),$ defined by the ansatz
\begin{equation}
\Phi = \frac{1}{\sqrt{2}}\,\left( \begin{array}{c} v+R+ i\,I_3 \\
-I_2 + i\,I_1
\end{array} \right)\ .                      \label{48}
\end{equation}
And, as in the abelian case, we replace the coupling constants
$\mu,\;\lambda,$ as parameters of the theory by
\begin{equation}
m = \frac{v\,g}{2}\ ,\qquad M = \sqrt{2}\,\mu\ ,  \label{49}
\end{equation}
which turn out to be the (unrenormalized) masses of the gauge
bosons and the Higgs particle respectively. The field equations of
the model look exactly like (\ref{15})--(\ref{17}), except that
there are now three $A_a^{\mu}$-equations and three
$I_a$-equations, one for each value of the group index $a$.
Correspondingly we get now three consistence conditions:
\begin{equation}
\mathcal{F}_a := \partial_{\mu}\mathcal{R}^{\mu}_a +
m\,\mathcal{R}_{Ia} = 0\ .                 \label{50}
\end{equation}

Wightman gauges can be defined and constructed like in the abelian
case. We are here not concerned with maximal generality, but need
only consider the U- and the R-gauge. The {\em U-gauge} can again
be characterized by the gauge condition $I_a=0$ for all $a$. Its
surviving cross propagators are taken over from (\ref{33}) as
\begin{equation}
w^{\mu\nu}_{ab}(p) = -\,\delta_{ab}\,(g^{\mu\nu}-
m^{-2}p^{\mu}p^{\nu})\:\delta^m_+(p)\ ,\quad w_{RR}(p) =
\delta^M_+(p)\ ,             \label{51}
\end{equation}
and similarly for the sector propagators. The {\em R-gauge} is
again defined by the propagators (\ref{41}), where the first three
lines hold for $a$-$a$ propagators for any value of the group
index $a$, while the mixed $a$-$b$ propagators with $a\neq b$
vanish.

The physical space $\mathcal{V}_{ph}$ is again equated with the
state space $\mathcal{V}_U$ of the U-gauge. In order to turn this
into a gauge invariant definition also usable in the R-gauge, we
must again produce $\mathcal{V}_U$ from the vacuum by applying
gauge invariant fields. As one of these fields we use the Higgs
field, which is now defined as
\begin{equation}
\Psi(x) = R(x) + \frac{g}{4m}\,\big[R^2(x) + I_a(x)\,I_a(x)\big]\
. \label{52}
\end{equation}
But the $F_a^{\alpha\beta}$ are no longer gauge invariant.
However, we can replace them by gauge invariant fields, which we
choose to be those introduced by Fr\"ohlich et al.\cite{FMS}. As
one of them we define
\begin{equation}
V_3^{\mu\nu}(x) :=
\frac{i\,g^2}{m^2}\,\Phi^*(x)\,T_a\,F_a^{\mu\nu}(x)\,\Phi(x)\ ,
\label{53}
\end{equation}
where $\Phi$ is expressed by (\ref{48}) with $v=2m/g$. In the
U-gauge this becomes
\[ V_3 = F_3 + \frac{g}{m}\,R\,F_3 + \frac{g^2}{4m^2}\,R^2\,F_3\ .     \]
$V_2^{\mu\nu}$ is defined in the same way, except that the $T_a$
are replaced by their cyclic permutation $(T_1\to T_2,\;T_2\to
T_3,\;T_3\to T_1)$. Repeating this operation we obtain
$V_1^{\mu\nu}$. By the same kind of arguments as used in Sect.~4
it can be shown that the restrictions to the U-gauge of these
$V_a$, together with $\Psi$, indeed reproduce $\mathcal{V}_U$.

 \medskip

Hence again, the only W-functions of direct physical relevance are
those containing only the physical fields $\Psi,\;V_a$, and only
the renormalizability of these must be decided. And this is again
easiest to achieve in the R-gauge. The method used is the same as
in the abelian case. It turns out to be more complicated in its
details. The main reason for this is that the simple form
(\ref{19}) of $\mathcal{F}_3$ is replaced by the more complicated
expression
\begin{eqnarray}
\mathcal{F}_{a3} &=&
g\,\varepsilon_{abc}\,A_{c\nu}\,\big[(\Box+m^2)\,A^{\nu}_b-
\partial^{\nu} \partial^{\mu}\!A_{b\mu} + m\,\partial^{\nu}I_b\big]
\nonumber \\ & & + \frac{g}{2}\,I_a\,(\Box+M^2)\,R -
\frac{g}{2}\,R\,
(\Box I_a + m\,\partial_{\mu}A^{\mu}_a) \nonumber \\
 & & + \frac{g}{2}\, \varepsilon_{abc}\,I_b\,(\Box I_c +
 m\,\partial_{\mu}A^{\mu}_c)\ .     \label{54}
\end{eqnarray}
Including this as a sum of composite external vertices in a sector
in which the mixed $A$--$I$ propagators are already eliminated, we
find that
\[ \partial_{\mu}A_a^{\mu} = 0\ ,   \]
so that the corresponding terms in (\ref{54}) can be dropped. But
even so the terms in the first line of (\ref{54}) do not have the
desired fusing effect on the adjacent propagators. The factor
$(p^2-m^2)$ of the first term applied to an
$A^{\nu}$-$A^{\lambda}$ propagator produces the ghost term
$p^{\nu}p^{\lambda}/(m^2\,p^2)$, and the $p^{\nu}I_b$ term applied
to a $I$-$I$ propagator clearly does not remove its singularity at
$p^2=0$. Hence, even if the fusing contributions do cancel like in
the abelian case, there remains a non-fusing contribution. But the
two offensive terms combine in such a way that they produce a
ghost line ending in a new $\mathcal{F}$ vertex, now inside the
sector, which fact allows using an inductive procedure leading to
a simple result. It turns out that the undesirable non-fusing
terms can be removed by the introduction of  Faddeev-Popov ghost
loops (FP loops)\cite{FP}. Such a loop is a directed closed loop.
Each line carries a propagator
\[ \frac{i}{2\,\pi\,(p^2+i\epsilon)}  \]
(in a $T^+$ sector) and a group index $a,\cdots$. The loop
contains only 3-line vertices with an $A_c^{\nu}$ line joining the
loop. The vertex factor is
\[ (2\sqrt{2\pi})^{-1}\,g\,\varepsilon_{abc}\,(p^{\nu}+q^{\nu})  \]
with $p$ the loop momentum leaving the vertex, $q$ that entering
the vertex, and $a$ and $b$ are the indices of the lines
respectively leaving and entering the vertex. And each such ghost
loop contributes an extra factor $-1$.

 \medskip

We might then {\em conjecture} the following generalization of the
Theorem of Sect.~5 to hold:

\begin{it}
Change the graph rules of the R-gauge by omitting the mixed
$A^{\mu}_a$-$I_a$ propagators and changing the signs of the
$I_a$-$I_a$ propagators, and by admitting an arbitrary number of
FP-loops. This procedure does not change the W-functions and
related functions of the physical fields $V_a^{\mu\nu},\;\Psi$.
\end{it}

 These conjectured rules are again
 the rules of the standard formalism in the Landau gauge.

The conjecture would be correct, if the fusing terms of
$\mathcal{F}_3$ did lead to graph-local cancellations in analogy
to the abelian case. This turns out to be the case for purely
internal cancellations, that is if the end points of the fused
lines are internal $L_{int}$ vertices. But it is  not true in all
cases where external vertices (composite fields contributing to
$V_a$) are involved. Therefore the equality of the physical
W-functions in the Landau gauge and the R-gauge, and hence in the
HKM in general, cannot be proved. This should not be interpreted
as a weakness of our method. There are strong indications that the
Landau gauge is indeed not  physically equivalent to the HKM, if
`physical equivalence' is defined in our sense, not simply as the
equality of the S-matrices.

As a result, there exists as yet no convincing proof of the full
renormalizability of the non-abelian HKM.


\begin{thebibliography}{1}
\bibitem{FP} L.~D.~Faddev, and V.~N.~Popov: {\it Phys. Letters 25B, 29
(1967)}.
\bibitem{FMS} J.~Fr\"ohlich, G.~Morchio, and F.~Strocchi: {\it Nucl. Phys. B190, 553
(1981)}.
\bibitem{IZ} C.~Itzykson, and J.-B.~Zuber. {\it Quantum Field
Theory.} McGraw-Hill, New York, 1980.
\bibitem{Os} A.~Ostendorf: {\it Ann.~Inst.~H.~Poincar\'e 40, 273
(1984)}.
\bibitem{St2} O.~Steinmann: {\it Commun.~Math.~Phys. 152, 627
(1993)}.
\bibitem{St1} O.~Steinmann: {\it Perturbative QED and Axiomatic Field
Theory.} Springer, Berlin, 2000.
\bibitem{SW} R.~F.~Streater, and A.~S.~Wightman: {\it PCT, Spin and Statistics, and All
That.} Benjamin/Cummings, Reading MA, 1978.
\bibitem{We} S.~Weinberg: {\it The Quantum Theory of Fields,
Vol.~2.} Cambridge U. Press, Cambridge, 1996.
\bibitem{Zi} W.~Zimmermann, in: {\it Lectures on Elementary Particles and Quantum Field Theory (ed. S.~Deser et al.).}
MIT Press, Cambridge MA, 1971.
\end{thebibliography}
\end{document}